\documentclass[twocolumn,preprintnumbers,pra,hyperref]{revtex4}
\usepackage{amssymb}
\usepackage{float}
\usepackage{graphics}
\usepackage{amsfonts}
\usepackage{amsmath}

\begin{document}

\title{Evolution of physical observables and entropy in laser process
studied on the basis of Kraus-form solution of laser's master equation}
\author{Junhua Chen$^{1,2,3}$ and Hongyi Fan$^{3}$\thanks{%
Email: fhym@ustc.edu.cn}}
\affiliation{$^{1}$ Hefei Center for Physical Science and Technology, Hefei, Anhui,
230026, China\\
$^{2}$ CAS Key Laboratory of Materials for Energy Conversion, Hefei, Anhui,
230026, China\\
$^{3}$ Department of Material Science and Engineering, USTC, Hefei, Anhui,
230026, China}

\begin{abstract}
Though laser physics began at 1960s, its entropy evolution has not been
touched until very recently the Kraus-form solution of laser's master
equation is reported (Ann. Phys. 334 (2013)). We study the new physics based
on the discovery in this paper. We analyze time evolution of physical
observables and quantum optical properties in the laser process with
arbitrary initial states, such as the photon number, the second degree of
coherence, etc. The evolution of entropy of these states is also studied.
Our results well conform with the known behaviour of laser, which confirms
that the master equation describes laser suitably, and the Kraus-form
operator solution is correct, elegant and useful.\newline
\end{abstract}

\maketitle

\section{Introduction}

The birth of laser, opening up the research area of quantum optics in 1960s,
is one of the greatest inventions of mankind. Quantum mechanically, the
stable laser above some threshold is described by a coherent state. However,
as a thermo object, the entropy evolution of a laser process has not been
touched until very recently we evaluated it after solving time evolution
master equation of the laser\textsuperscript{\cite{a1,a2,a3,a4}}%
\begin{eqnarray}
\frac{d\rho (t)}{dt} &=&g[2a^{\dag }\rho (t)a-aa^{\dag }\rho (t)-\rho
(t)aa^{\dag }]  \label{1} \\
&&+\kappa \lbrack 2a\rho (t)a^{\dag }-a^{\dag }a\rho (t)-\rho (t)a^{\dag }a],
\notag
\end{eqnarray}%
and obtained its solution in the form of infinite sum (or named Kraus form
solution)%
\begin{equation}
\rho (t)=\sum\limits_{i,j=0}^{\infty }M_{ij}\rho _{0}M_{ij}^{\dag }.
\label{2}
\end{equation}%
$M_{ij}$ is usually called Kraus operator\textsuperscript{\cite{a5,a6}}. In
Eqn. (\ref{1}) $g$ and $\kappa $ represent the cavity gain and loss
respectively, $a^{\dag }$ and $a$ are photon creation and annihilation
operator. For Eqn. (\ref{2}) we have used the entangled state representation%
\textsuperscript{\cite{a7}} to derive 
\begin{equation}
M_{ij}=\sqrt{\frac{T_{3}\kappa ^{i}g^{j}T_{1}^{i+j}}{i!j!T_{2}^{2j}}}%
e^{a^{\dag }a\ln T_{2}}a^{\dag j}a^{i},  \label{3}
\end{equation}%
and 
\begin{eqnarray}
T_{1} &=&\frac{1-e^{-2(\kappa -g)t}}{\kappa -ge^{-2(\kappa -g)t}},\qquad
T_{2}=\frac{(\kappa -g)e^{-(\kappa -g)t}}{\kappa -ge^{-2(\kappa -g)t}},
\label{4} \\
\qquad T_{3} &=&\frac{\kappa -g}{\kappa -ge^{-2(\kappa -g)t}}.  \notag
\end{eqnarray}%
It is noticeable that they are not independent of each other, in fact 
\begin{equation}
T_{3}=1-gT_{1},\text{ }T_{2}^{2}/T_{3}=1-\kappa T_{1}.  \label{5}
\end{equation}%
One can check the trace conservative law 
\begin{equation}
\sum\limits_{i,j=0}^{\infty }M_{ij}^{\dag }M_{ij}=1.  \label{6}
\end{equation}

Since laser's evolution and properties are very important we must answer the
important question: does the solution in Eqs. (\ref{2}-\ref{4}) can indeed
reflect laser channel's physical properties? In this work, we shall study
the evolution of some important physical quantities in laser process such as
the photon number, the second degree of coherence, for laser with arbitrary
initial states. Since entropy involves both thermodynamic and informative
properties, it is important to explore the evolution of entropy of these
states. As one can see shortly later, our results well conform with the
behavior of laser, which confirms that our solution Eqs. (\ref{2}-\ref{4})
is correct, elegant and useful.

\section{The Evolution of $\protect\rho $}

In terms of the number state $\left\vert m\right\rangle =\frac{a^{\dagger m}%
}{\sqrt{m!}}\left\vert 0\right\rangle $ in Fock space, and 
\begin{equation*}
a\left\vert m\right\rangle =\sqrt{m}\left\vert m-1\right\rangle ,\text{ }%
a^{\dagger }\left\vert m\right\rangle =\sqrt{m+1}\left\vert m+1\right\rangle
.
\end{equation*}%
For arbitrary initial state $\rho _{0}=\sum\limits_{m,n=0}^{\infty }\rho
_{m,n}\left\vert m\right\rangle \left\langle n\right\vert $, using Eqs. (\ref%
{2}, \ref{3}) we have%
\begin{equation}
\begin{array}{c}
\rho (t)=\sum\limits_{i,j=0}^{\infty }M_{ij}\rho _{0}M_{ij}^{\dag } \\ 
=\sum\limits_{i,j=0}^{\infty }\sum\limits_{m,n=0}^{\infty }\frac{T_{3}\kappa
^{i}g^{j}T_{1}^{i+j}T_{2}^{m+n-2i}}{i!j!}\sqrt{\frac{m!n!\left( m-i+j\right)
!\left( n-i+j\right) !}{\left( m-i\right) !^{2}\left( n-i\right) !^{2}}} \\ 
\times \rho _{mn}\left\vert m-i+j\right\rangle \left\langle n-i+j\right\vert
\\ 
=\sum\limits_{i,j=0}^{\infty }\sum\limits_{m^{\prime },n^{\prime
}=0}^{\infty }\frac{T_{3}\kappa ^{i}g^{j}T_{1}^{i+j}T_{2}^{m^{\prime
}+n^{\prime }}}{i!j!}\sqrt{\frac{\left( m^{\prime }+i\right) !\left(
n^{\prime }+i\right) !\left( m\prime +j\right) !\left( n\prime +j\right) !}{%
m^{\prime }!^{2}n^{\prime }!^{2}}} \\ 
\times \rho _{m^{\prime }+i,n^{\prime }+i}\left\vert m^{\prime
}+j\right\rangle \left\langle n^{\prime }+j\right\vert .%
\end{array}
\label{7}
\end{equation}%
We now discuss $\rho (t)$ for different working conditions of laser as
follows.

If $\kappa >g$, gain is less than loss, then from Eqn. (\ref{4}) we know 
\begin{eqnarray}
T_{1} &=&\frac{1}{\kappa }+O\left( e^{-2\left( \kappa -g\right) t}\right) ,\ 
\label{8} \\
T_{2} &=&\frac{\kappa -g}{\kappa }e^{-\left( \kappa -g\right) t}+O\left(
e^{-3\left( \kappa -g\right) t}\right) ,  \notag \\
T_{3} &=&\frac{\kappa -g}{\kappa }+O\left( e^{-2\left( \kappa -g\right)
t}\right) .  \notag
\end{eqnarray}%
Therefore only those terms with $m^{\prime }=n^{\prime }=0$ in Eqn. (\ref{7}%
) will contribute to $\rho (+\infty )$ (long time limit) 
\begin{eqnarray}
\rho (+\infty ) &=&\sum\limits_{i,j=0}^{\infty }T_{3}\left( +\infty \right)
\kappa ^{i}g^{j}T_{1}^{i+j}\left( +\infty \right) \rho _{ii}\left\vert
j\right\rangle \left\langle j\right\vert  \label{9} \\
&=&\frac{\kappa -g}{\kappa }\sum\limits_{j=0}^{\infty }\left( \frac{g}{%
\kappa }\right) ^{j}\left\vert j\right\rangle \left\langle j\right\vert
\sum\limits_{i=0}^{\infty }\rho _{ii}  \notag \\
&=&\left( 1-e^{-\ln \frac{\kappa }{g}}\right) \sum\limits_{j=0}^{\infty
}e^{-j\ln \frac{\kappa }{g}}\left\vert j\right\rangle \left\langle
j\right\vert .  \notag
\end{eqnarray}%
The quantum state approaches to therm-equilibrium state with equivalent
temperature $T=\frac{\hbar \omega }{k_{B}\ln \frac{\kappa }{g}}$, where $%
k_{B}$ is the Boltzmann constant. The properties of therm-equilibrium state
are well-known to the physicists.

If $\kappa <g$, then 
\begin{eqnarray}
T_{1} &=&\frac{1}{g}-\frac{g-\kappa }{g^{2}}e^{2\left( \kappa -g\right)
t}+O\left( e^{3\left( \kappa -g\right) t}\right) ,  \label{10} \\
T_{2} &=&\frac{g-\kappa }{g}e^{\left( \kappa -g\right) t}+O\left( e^{3\left(
\kappa -g\right) t}\right) ,  \notag \\
T_{3} &=&\frac{g-\kappa }{g}e^{2\left( \kappa -g\right) t}+O\left(
e^{3\left( \kappa -g\right) t}\right) .  \notag
\end{eqnarray}%
The overall factor $T_{3}$ of $\rho (t)$\ in Eqn. (\ref{7}) is also
exponentially small, therefore $\rho (t)$ will not approach to any specific
state when $t\rightarrow +\infty $. More works are need to analysis the
behavior of the laser when $\kappa <g$.

\section{Evolution of the Expectation of Physical Quantities in Laser}

The expectation of a physical quantity $\hat{A}$ for a system described by
density operator $\rho $ is defined by $\left\langle \hat{A}\right\rangle
\equiv Tr\left( \hat{A}\rho \right) $. Since $\rho $ evolves with time, $%
\left\langle \hat{A}\right\rangle $ is a function of time too. For laser
processes with different initial states $\rho _{0}$, it will be overwhelming
to calculate $\rho \left( t\right) =\sum_{i,j=0}^{\infty }M_{ij}\rho
_{0}M_{ij}^{\dagger }$ for each $\rho _{0}$. Noticing that 
\begin{eqnarray}
\left\langle \hat{A}\right\rangle _{t} &\equiv &Tr\left[ \hat{A}\rho \left(
t\right) \right] =Tr\left( \sum_{i,j=0}^{\infty }\hat{A}M_{ij}\rho
_{0}M_{ij}^{\dagger }\right)  \label{11} \\
&=&Tr\left( \sum_{i,j=0}^{\infty }M_{ij}^{\dagger }\hat{A}M_{ij}\rho
_{0}\right) =Tr\left( \hat{A}_{t}\rho _{0}\right) ,  \notag
\end{eqnarray}%
where the evolving $\hat{A}$ operator is defined as%
\begin{eqnarray}
\hat{A}_{t} &\equiv &\sum_{i,j=0}^{\infty }M_{ij}^{\dagger }\hat{A}M_{ij}
\label{12} \\
&=&\sum_{i,j=0}^{\infty }\frac{T_{3}\kappa ^{i}g^{j}T_{1}^{i+j}}{%
i!j!T_{2}^{2j}}a^{\dag i}a^{j}e^{a^{\dag }a\ln T_{2}}\hat{A}e^{a^{\dag }a\ln
T_{2}}a^{\dag j}a^{i}.  \notag
\end{eqnarray}%
Thus instead of calculating $\rho \left( t\right) $ for each $\rho _{0}$, we
just need to calculate the evolving $\hat{A}$ operator $\hat{A}_{t}$ for
each $\hat{A}$. Once we obtain $\hat{A}_{t}$, we can calculate $\left\langle 
\hat{A}\right\rangle _{t}\equiv Tr\left( \hat{A}_{t}\rho _{0}\right) $ for
arbitrary initial state $\rho _{0}$ straightforwardly. Further, even when
the operator $\hat{A}\left( t\right) $ itself is time dependent, we can
still define the evolving $\hat{A}\left( t\right) $ operator $\hat{A}\left(
t\right) _{t}\equiv \sum_{i,j=0}^{\infty }M_{ij}^{\dagger }\hat{A}\left(
t\right) M_{ij}$. And we still have $\left\langle \hat{A}\left( t\right)
\right\rangle _{t}\equiv Tr\left[ \hat{A}\left( t\right) \rho \left(
t\right) \right] =Tr\left( \hat{A}\left( t\right) _{t}\rho _{0}\right) $.

\section{The Generating Function of the Evolving $\left( a^{\dagger
}a\right) ^{m}$ Operators}

In this section we do some preparatory work for calculating the evolution of
the expected photon number and the second degree of coherence in lasers.

Since%
\begin{equation}
\sum\limits_{m=0}^{\infty }\frac{1}{m!}\lambda ^{m}\left( a^{\dagger
}a\right) ^{m}=e^{\lambda a^{\dagger }a},  \label{13}
\end{equation}%
we have%
\begin{equation}
\sum\limits_{m=0}^{\infty }\frac{1}{m!}\lambda ^{m}\left[ \left( a^{\dagger
}a\right) ^{m}\right] _{t}=\left( e^{\lambda a^{\dagger }a}\right) _{t}.
\label{14}
\end{equation}%
Once we obtain $\left( e^{\lambda a^{\dagger }a}\right) _{t}$, the evolving
photon number operators $\left[ \left( a^{\dagger }a\right) ^{m}\right] _{t}$
can be calculated using the generating function $\left( e^{\lambda
a^{\dagger }a}\right) _{t}$ (this is also named the cumulant expansion): 
\begin{equation}
\left[ \left( a^{\dagger }a\right) ^{m}\right] _{t}=\frac{d^{m}}{d\lambda
^{m}}\left( e^{\lambda a^{\dagger }a}\right) _{t}|_{\lambda =0}.  \label{15}
\end{equation}%
Similarly, since 
\begin{equation}
\sum\limits_{m=0}^{\infty }\frac{1}{m!}\mu ^{m}a^{\dagger m}a^{m}=:e^{\mu
a^{\dagger }a}:=\exp \left[ \ln \left( 1+\mu \right) a^{\dagger }a\right] ,
\label{16}
\end{equation}%
we have%
\begin{equation}
\left( a^{\dagger m}a^{m}\right) _{t}=\frac{d^{m}}{d\mu ^{m}}\left( e^{\ln
\left( 1+\mu \right) a^{\dagger }a}\right) _{t}|_{\mu =0}.  \label{17}
\end{equation}%
Now we employ the completeness relation of the coherent state $\left\vert
z\right\rangle =\exp \left[ -\frac{|z|^{2}}{2}+za^{\dagger }\right]
\left\vert 0\right\rangle $ 
\begin{equation}
\int \frac{d^{2}z}{\pi }\left\vert z\right\rangle \left\langle z\right\vert
=\int \frac{d^{2}z}{\pi }:e^{-\left\vert z\right\vert ^{2}+a^{\dagger
}z+z^{\ast }a-a^{\dagger }a}:=1  \label{18}
\end{equation}%
and the method of integration within an ordered product of operators%
\textsuperscript{\cite{a8,a9}} to calculate the generating function $\left(
e^{\lambda a^{\dagger }a}\right) _{t}$%
\begin{equation}
\begin{array}{c}
\left( e^{\lambda a^{\dagger }a}\right) _{t}=\sum_{i,j=0}^{\infty }\frac{%
T_{3}\kappa ^{i}g^{j}T_{1}^{i+j}}{i!j!T_{2}^{2j}}a^{\dagger i}a^{j}e^{\left(
2\ln T_{2}+\lambda \right) a^{\dagger }a}a^{\dagger j}a^{i} \\ 
=\sum_{i,j=0}^{\infty }\frac{T_{3}\kappa ^{i}g^{j}T_{1}^{i+j}}{i!j!}%
e^{j\lambda }a^{\dagger i}e^{\left( \lambda +2\ln T_{2}\right) a^{\dagger
}a}a^{j}a^{\dagger j}a^{i} \\ 
=\int \frac{d^{2}z}{\pi }\sum_{i,j=0}^{\infty }\frac{T_{3}\kappa
^{i}g^{j}T_{1}^{i+j}}{i!j!}e^{j\lambda }a^{\dagger i}:e^{\left(
T_{2}^{2}e^{\lambda }-1\right) a^{\dagger }a}:z^{j}\left\vert z\right\rangle
\left\langle z\right\vert z^{\ast j}a^{i} \\ 
=T_{3}\int \frac{d^{2}z}{\pi }:e^{T_{2}^{2}e^{\lambda }a^{\dagger }z+\left(
gT_{1}e^{\lambda }-1\right) \left\vert z\right\vert ^{2}+\left( \kappa
T_{1}-1\right) a^{\dagger }a+z^{\ast }a}: \\ 
=\frac{T_{3}}{1-gT_{1}e^{\lambda }}:\exp \left[ \left( \frac{%
T_{2}^{2}e^{\lambda }}{1-gT_{1}e^{\lambda }}+\kappa T_{1}-1\right)
a^{\dagger }a\right] : \\ 
=\frac{T_{3}}{1-gT_{1}e^{\lambda }}\exp \left[ a^{\dagger }a\ln \left( \frac{%
T_{2}^{2}e^{\lambda }}{1-gT_{1}e^{\lambda }}+\kappa T_{1}\right) \right] .%
\end{array}
\label{19}
\end{equation}%
Note that the convergence of the integration over $z$ demands that $%
1-gT_{1}e^{\lambda }>0$, $\lambda <-\ln \left( gT_{1}\right) $.

We can immediately write down%
\begin{equation}
\begin{array}{c}
\left( :e^{\mu a^{\dagger }a}:\right) _{t}=\left( e^{\ln \left( 1+\mu
\right) a^{\dagger }a}\right) _{t} \\ 
=\frac{T_{3}}{1-gT_{1}\left( 1+\mu \right) }:\exp \left[ \left( \frac{%
T_{2}^{2}\left( 1+\mu \right) }{1-gT_{1}\left( 1+\mu \right) }+\kappa
T_{1}-1\right) a^{\dagger }a\right] :%
\end{array}
\label{20}
\end{equation}%
Thus using Eqs. (\ref{17}, \ref{20}) we have 
\begin{equation}
\left( a^{\dagger }a\right) _{t}=e^{2(g-\kappa )t}\left( a^{\dagger }a+\frac{%
g}{g-\kappa }\right) -\frac{g}{g-\kappa }  \label{21}
\end{equation}%
and%
\begin{equation}
\left( a^{\dagger 2}a^{2}\right) _{t}=\frac{2g^{2}T_{1}^{2}}{T_{3}^{2}}+4%
\frac{gT_{1}}{T_{3}}\frac{T_{2}^{2}}{T_{3}^{2}}a^{\dagger }a+\frac{T_{2}^{4}%
}{T_{3}^{4}}a^{\dagger 2}a^{2}.  \label{22}
\end{equation}%
The photon number expectation at time $t$ is%
\begin{equation}
\left\langle a^{\dagger }a\right\rangle _{t}=e^{2\left( g-\kappa \right)
t}\left( \left\langle n\right\rangle _{0}+\frac{g}{g-\kappa }\right) -\frac{g%
}{g-\kappa }.  \label{23}
\end{equation}%
And if $g>\kappa $ we have further 
\begin{equation}
\left\langle \left( a^{\dagger }a\right) ^{\delta }\right\rangle \sim
N_{\delta }e^{2\delta \left( g-\kappa \right) t}  \label{24}
\end{equation}%
as $t\rightarrow +\infty $ , where $N_{\delta }$ is a linear combination of $%
\left\langle \left( a^{\dagger }a\right) ^{\delta ^{\prime }}\right\rangle
_{0}$ ($\delta ^{\prime }\leqslant \delta $).

We have the following conclusions for arbitrary initial states:

If $g<\kappa $, damping is larger than pumping, then $\left\langle
a^{\dagger }a\right\rangle _{t}\sim \frac{g}{\kappa -g}$ as $t\rightarrow
+\infty $, as one should expect for therm-equilibrium state with temperature 
$T=\frac{\hbar \omega }{k_{B}\ln \frac{\kappa }{g}}$.

If $g>\kappa $, pumping is larger than damping, then for any initial states
the system's photon number $\left\langle a^{\dagger }a\right\rangle _{t}$
will increase exponentially, $\left\langle a^{\dagger }a\right\rangle
_{t}\sim e^{2\left( g-\kappa \right) t}\left( \left\langle \hat{n}%
\right\rangle _{0}+\frac{g}{g-\kappa }\right) $, as a laser should behave
when it is working well.

Now we examine how the second degree of coherence $\mathfrak{g}_{0}^{\left(
2\right) }=\frac{\left\langle a^{\dagger 2}a^{2}\right\rangle _{0}}{%
\left\langle a^{\dagger }a\right\rangle _{0}^{2}}$ evolves into $\mathfrak{g}%
^{\left( 2\right) }\left( t\right) =\frac{\left\langle a^{\dagger
2}a^{2}\right\rangle _{t}}{\left\langle a^{\dagger }a\right\rangle _{t}^{2}}$%
. According to Eqs. (\ref{21}, \ref{22}) we have%
\begin{equation}
\mathfrak{g}^{\left( 2\right) }\left( t\right) =\frac{\left\langle
a^{\dagger 2}a^{2}\right\rangle _{t}}{\left\langle a^{\dagger
}a\right\rangle _{t}^{2}}=2+\frac{\mathfrak{g}_{0}^{\left( 2\right) }-2}{%
\left( 1+\chi \right) ^{2}}  \label{25}
\end{equation}%
where%
\begin{equation}
\chi \left( t\right) =\frac{gT_{1}}{e^{2\left( g-\kappa \right)
t}\left\langle \hat{n}\right\rangle _{0}T_{3}}.  \label{26}
\end{equation}%
When $g\leqslant \kappa $, $\chi \left( t\right) =\frac{gT_{1}}{e^{2\left(
g-\kappa \right) t}\left\langle n\right\rangle _{0}T_{3}}\rightarrow +\infty 
$, therefore $g^{\left( 2\right) }\left( +\infty \right) \equiv 2$ for any
initial state $\rho _{0}$.

When $g>\kappa $, $\chi \left( +\infty \right) =\frac{g}{\left( g-\kappa
\right) \left\langle n\right\rangle _{0}}>0$,%
\begin{equation}
\mathfrak{g}^{\left( 2\right) }\left( +\infty \right) =2+\frac{\mathfrak{g}%
_{0}^{\left( 2\right) }-2}{\left( 1+\chi \left( +\infty \right) \right) ^{2}}%
.  \label{27}
\end{equation}%
For arbitrary initial state $\rho _{0}$, we have 
\begin{equation}
\mathfrak{g}_{0}^{\left( 2\right) }=\frac{\left\langle \hat{n}\right\rangle
_{0}-1}{\left\langle \hat{n}\right\rangle _{0}}+\frac{\left\langle \hat{n}%
^{2}\right\rangle _{0}-\left\langle \hat{n}\right\rangle _{0}^{2}}{%
\left\langle \hat{n}\right\rangle _{0}^{2}}\geqslant 1-\frac{1}{\left\langle
n\right\rangle _{0}},  \label{28}
\end{equation}%
therefore%
\begin{eqnarray}
\mathfrak{g}^{\left( 2\right) }\left( +\infty \right) &=&2+\frac{%
g_{0}^{\left( 2\right) }-2}{\left( 1+\frac{g}{\left( g-\kappa \right)
\left\langle n\right\rangle _{0}}\right) ^{2}}  \label{29} \\
&\geqslant &2-\frac{1+\frac{1}{\left\langle n\right\rangle _{0}}}{\left( 1+%
\frac{g}{\left( g-\kappa \right) \left\langle n\right\rangle _{0}}\right)
^{2}}  \notag \\
&>&2-\frac{1+\frac{1}{\left\langle n\right\rangle _{0}}}{1+\frac{g}{\left(
g-\kappa \right) \left\langle n\right\rangle _{0}}}>1,  \notag
\end{eqnarray}%
i.e., when a laser works in $g>\kappa $ region, the photons of the laser
tend to be bunching.

\section{The Evolution of Entropy}

In our preceding paper\textsuperscript{\cite{a5}} for an initial state $\rho
_{0}=\left\vert z\right\rangle \left\langle z\right\vert $ we have derived
the exact entropy expression 
\begin{equation}
S\left( \rho _{z}\left( t\right) \right) =-k_{B}\left( \ln T_{3}+\frac{gT_{1}%
}{1-gT_{1}}\ln gT_{1}\right) ,  \label{30}
\end{equation}%
which is independent of the initial value $z$. Therefore for initial state $%
\rho _{0}=\int \frac{d^{2}z}{\pi }P\left( z\right) \left\vert z\right\rangle
\left\langle z\right\vert $ with positive coefficient in its
Glauber--Sudarshan P-representation (such quantum system has a classical
analog), by the concaveness of von Neumann entropy, we have the following
inequality%
\begin{eqnarray}
S\left( \rho \left( t\right) \right) &=&S\left[ \int \frac{d^{2}z}{\pi }%
P\left( z\right) \rho _{z}\left( t\right) \right]  \label{31} \\
&\geqslant &\int \frac{d^{2}z}{\pi }P\left( z\right) S\left[ \rho _{z}\left(
t\right) \right]  \notag \\
&=&S\left[ \rho _{z}\left( t\right) \right] \int \frac{d^{2}z}{\pi }P\left(
z\right)  \notag \\
&=&-k_{B}\left( \ln T_{3}+\frac{gT_{1}}{1-gT_{1}}\ln gT_{1}\right) ,  \notag
\end{eqnarray}%
which provides an estimation on the entropy.

Now we examine the evolution of the entropy for more general initial states.

According to Eqn. (\ref{22}), when $\kappa >g$, the density operator will
approach to $\rho (+\infty )=\frac{\kappa -g}{\kappa }\sum\limits_{j=0}^{%
\infty }\left( \frac{g}{\kappa }\right) ^{j}\left\vert j\right\rangle
\left\langle j\right\vert $ for arbitrary initial state $\rho
_{0}=\sum\limits_{m,n=0}^{\infty }\rho _{m,n}\left\vert m\right\rangle
\left\langle n\right\vert $. The entropy will approach to%
\begin{eqnarray}
S_{\infty } &=&-k_{B}\sum\limits_{j=0}^{\infty }\frac{\kappa -g}{\kappa }%
\left( \frac{g}{\kappa }\right) ^{j}\ln \left[ \frac{\kappa -g}{\kappa }%
\left( \frac{g}{\kappa }\right) ^{j}\right]  \label{32} \\
&=&k_{B}\left( \frac{g}{\kappa -g}\ln \frac{g}{\kappa }+\ln \frac{\kappa -g}{%
\kappa }\right) .  \notag
\end{eqnarray}

In the case $\kappa <g$, we first consider initial states which are diagonal
in photon-number representation, $\rho _{0}=\sum\limits_{m=0}^{\infty }\rho
_{m,m}\left\vert m\right\rangle \left\langle m\right\vert $. We have 
\begin{equation}
\rho (t)=\sum\limits_{i,j=0}^{\infty }M_{ij}\rho _{0}M_{ij}^{\dag
}=\sum\limits_{k=0}^{\infty }\rho _{kk}\left( t\right) \left\vert
k\right\rangle \left\langle k\right\vert ,  \label{33}
\end{equation}%
where%
\begin{equation}
\rho _{kk}\left( t\right) =\sum\limits_{m=0}^{k}\frac{T_{3}\left(
gT_{1}\right) ^{k-m}T_{2}^{2m}}{\left( k-m\right) !}\frac{k!}{m!^{2}}%
f^{\left( m\right) }\left( \kappa T_{1}\right)   \label{34}
\end{equation}%
and%
\begin{eqnarray}
f\left( x\right)  &\equiv &\sum\limits_{i=0}^{\infty }\rho _{ii}x^{i},
\label{35} \\
f^{\left( m\right) }\left( x\right)  &\equiv &\sum\limits_{i=0}^{\infty
}\rho _{m+i,m+i}\frac{\left( m+i\right) !}{i!}x^{i}=\frac{d^{m}}{dx^{m}}%
f\left( x\right) .  \notag
\end{eqnarray}%
Correspondingly, the von Neumann entropy is%
\begin{equation}
-S/k_{B}=\sum\limits_{k=0}^{\infty }\rho _{kk}\left( t\right) \ln \rho
_{kk}\left( t\right) =I_{1}+I_{2},  \label{36}
\end{equation}%
where%
\begin{eqnarray}
I_{1} &=&\sum\limits_{k=0}^{\infty }\rho _{kk}\left( t\right) \ln \left[
T_{3}\left( gT_{1}\right) ^{k}\right] ,  \label{36a} \\
I_{2} &=&\sum\limits_{k=0}^{\infty }\rho _{kk}\left( t\right) \ln \left[
\sum\limits_{m=0}^{k}\left( \frac{T_{2}^{2}}{gT_{1}}\right) ^{m}\frac{%
k!f^{\left( m\right) }\left( \kappa T_{1}\right) }{m!^{2}\left( k-m\right) !}%
\right] .  \notag
\end{eqnarray}%
The first term in Eqn. (\ref{36}) is%
\begin{equation}
\begin{array}{c}
I_{1}=\sum\limits_{k=0}^{\infty }\rho _{kk}\left( t\right) \ln
T_{3}+\sum\limits_{k=0}^{\infty }k\rho _{kk}\left( t\right) \ln gT_{1} \\ 
=\ln T_{3}+\left\langle \hat{n}\right\rangle _{t}\ln gT_{1} \\ 
=2\left( \kappa -g\right) t+\ln \frac{g-\kappa }{g}-\left( 1+\frac{g-\kappa 
}{g}\left\langle \hat{n}\right\rangle _{0}\right) +o\left( 1\right) 
\end{array}
\label{37}
\end{equation}%
when $t\rightarrow +\infty $. And $I_{2}$ is bounded below,%
\begin{equation}
I_{2}\geqslant \sum\limits_{k=0}^{\infty }\rho _{kk}\left( t\right) \ln
f\left( \kappa T_{1}\right) =\ln f\left( \kappa T_{1}\right) .  \label{37a}
\end{equation}

Noticing that $\rho _{ii}\geqslant 0$ and $f\left( 1\right)
=\sum\limits_{i=0}^{\infty }\rho _{ii}=1<+\infty $, we see that the radius
of convergence $R\geqslant 1$ for power series $f\left( x\right) \equiv
\sum\limits_{i=0}^{\infty }\rho _{ii}x^{i}$. Then positive power series $%
\sum\limits_{m=0}^{\infty }f^{\left( m\right) }\left( \kappa T_{1}\right) 
\frac{y^{m}}{m!}=f\left( \kappa T_{1}+y\right) $ converges for $0<y\leqslant
R-\kappa T_{1}$. Therefore each term $f^{\left( m\right) }\left( \kappa
T_{1}\right) \frac{y^{m}}{m!}\leqslant f\left( \kappa T_{1}+y\right) $ is
bounded with respect to $m$. We have%
\begin{eqnarray}
&&\sum\limits_{m=0}^{k}\left( \frac{T_{2}^{2}}{gT_{1}}\right) ^{m}\frac{%
k!f^{\left( m\right) }\left( \kappa T_{1}\right) }{m!^{2}\left( k-m\right) !}
\label{38} \\
&\leqslant &\sum\limits_{m=0}^{k}\left( \frac{T_{2}^{2}}{ygT_{1}}\right) ^{m}%
\frac{k!f\left( \kappa T_{1}+y\right) }{m!\left( k-m\right) !}  \notag \\
&=&f\left( \kappa T_{1}+y\right) \left( 1+\frac{T_{2}^{2}}{ygT_{1}}\right)
^{k},  \notag
\end{eqnarray}%
therefore the second term in Eqn. (\ref{36})%
\begin{eqnarray}
I_{2} &\leqslant &\sum\limits_{k=0}^{\infty }\rho _{kk}\left( t\right) \ln 
\left[ f\left( \kappa T_{1}+y\right) \left( 1+\frac{T_{2}^{2}}{ygT_{1}}%
\right) ^{k}\right]   \label{39} \\
&=&\ln f\left( \kappa T_{1}+y\right) +\left\langle \hat{n}\right\rangle
_{t}\ln \left( 1+\frac{T_{2}^{2}}{ygT_{1}}\right)   \notag \\
&\sim &\ln f\left( \frac{\kappa }{g}+y\right) +\left( \frac{g-\kappa }{g}%
\left\langle \hat{n}\right\rangle _{0}+1\right) \frac{g-\kappa }{gy}  \notag
\end{eqnarray}%
as $t\rightarrow +\infty $, which is a finite number for $0<y\leqslant R-%
\frac{\kappa }{g}$. Particularly, choosing $y=1-\frac{\kappa }{g}$ in Eqn. (%
\ref{39}), we see that 
\begin{equation}
I_{2}\lesssim \frac{g-\kappa }{g}\left\langle \hat{n}\right\rangle _{0}+1.
\label{39a}
\end{equation}%
Combining Eqn. (\ref{37a}, \ref{39a}), we see $\ln f\left( \kappa
T_{1}\right) \leqslant I_{2}\leqslant \frac{g-\kappa }{g}\left\langle \hat{n}%
\right\rangle _{0}+1$ is a finite number as $t\rightarrow +\infty $.

Finally from Eqs. (\ref{36}, \ref{37a}, \ref{39a}) we have%
\begin{equation}
\begin{array}{c}
2\left( g-\kappa \right) t+\ln \frac{g}{\left( g-\kappa \right) f\left(
\kappa /g\right) }+1+\frac{g-\kappa }{g}\left\langle \hat{n}\right\rangle
_{0} \\ 
\gtrsim S/k_{B}=2\left( g-\kappa \right) t+O\left( 1\right) \\ 
\gtrsim 2\left( g-\kappa \right) t+\ln \frac{g}{g-\kappa }%
\end{array}
\label{40}
\end{equation}%
as $t\rightarrow +\infty $ for laser with initial state $\rho
_{0}=\sum\limits_{m=0}^{\infty }\rho _{m,m}\left\vert m\right\rangle
\left\langle m\right\vert $ and $g>\kappa $. The specific entropy 
\begin{equation*}
\frac{S}{\langle n\rangle }\sim \frac{2k_{B}\left( g-\kappa \right) t}{\frac{%
g}{g-\kappa }+\langle n\rangle _{0}}e^{-2(g-\kappa )t}
\end{equation*}%
goes to zero exponentially.

For arbitrary initial states $\rho _{0}=\sum\limits_{m,n=0}^{\infty }\rho
_{m,n}\left\vert m\right\rangle \left\langle n\right\vert $, we see from
Eqn. (\ref{7}) that the off-diagonal elements of $\rho \left( t\right) $ are
exponentially small compared with diagonal elements as $t\rightarrow +\infty 
$\ when $g>\kappa $, i.e., $\rho \left( t\right) $ tends to be diagonal for
laser with arbitrary initial state. So we can well assume that $%
S=2k_{B}\left( g-\kappa \right) t+O\left( 1\right) $ as $t\rightarrow
+\infty $ for laser with arbitrary initial state when $g>\kappa $.

This result affirms that when a laser is working properly ($g>\kappa $), the
entropy increases linearly with time, yet the expected number of photons
increases much faster, therefore the specific entropy will goes to zero
exponentially. The photons in the laser are highly coherent and bunching in
this case.

\section{Summary}

In this paper, we analyze the laser process with arbitrary initial states,
and obtain the evolution law of the photon number, the second degree of
coherence and the entropy. If $\kappa >g$, then the photons in the laser
will approach a therm-equilibrium state with equivalent temperature $T=\frac{%
\hbar \omega }{k_{B}\ln \frac{\kappa }{g}}$. The expected photon number
approaches to $\frac{g}{\kappa -g}$, the second degree of coherence $%
g^{\left( 2\right) }$ approaches to $2$. The entropy approaches$\ $to $%
k_{B}\left( \frac{g}{\kappa -g}\ln \frac{\kappa }{g}+\ln \frac{\kappa }{%
\kappa -g}\right) $. If $g>\kappa $, then the photon number will increase
exponentially, $\left\langle n\right\rangle =e^{2\left( g-\kappa \right)
t}\left( \left\langle n\right\rangle _{0}+\frac{g}{g-\kappa }\right) -\frac{g%
}{g-\kappa }$, and the second degree of coherence $g^{\left( 2\right) }$
approaches to $2+\frac{g_{0}^{\left( 2\right) }-2}{\left( 1+\frac{g}{\left(
g-\kappa \right) \left\langle n\right\rangle _{0}}\right) ^{2}}>1$. For $%
\rho _{0}=\sum\limits_{m=0}^{\infty }\rho _{m,m}\left\vert m\right\rangle
\left\langle m\right\vert $ in $g>\kappa $ case, we proved that the entropy
will increase linearly, $S\sim 2k_{B}\left( g-\kappa \right) t$. All these
results conform with the known behavior of laser, this confirms that the
master equation (\ref{1}) describes laser's behavior well, and the
Kraus-form operator solution (\ref{2})-(\ref{4}) is correct, elegant and
useful.

\textbf{Acknowledgments: }Work supported by the National Natural Science
Foundation of China under grant 11105133 and 11751113, and the National
Basic Research Program of China (973 Program, 2012CB922001).

\end{document}